\documentclass[11pt,a4paper]{article}
\pdfoutput=1 

\usepackage[table]{xcolor}
\usepackage{colortbl}
\RequirePackage{ifpdf} 
\usepackage{amsmath} 
\usepackage{mathtools}

\usepackage{jheppub}
\usepackage{pstricks}
\usepackage[final]{pdfpages} 
\usepackage{ifpdf} 
\usepackage{slashed}

\usepackage[normalem]{ulem}
\usepackage{color}
\usepackage{xcolor}
\definecolor{urlblue}{rgb}{0.2,0.4,0.7}
\definecolor{citegreen}{rgb}{0,0.6,0.2}
\definecolor{linkred}{rgb}{0.9,0.2,0.1}
\usepackage{hyperref}
\hypersetup{
colorlinks=true, citecolor=citegreen, linkcolor=blue, urlcolor=urlblue}

\usepackage{graphics}
\usepackage{etoolbox} 
\usepackage{fixmath}
\usepackage{psfrag}

\usepackage{notoccite} 

\usepackage{amsfonts}
\usepackage{autobreak}
\usepackage{marginnote}
\usepackage{enumitem}
\usepackage{appendix}
\usepackage{caption} 
\newcommand{\NOdisplay}[1]{ }

\def\MSbar{\overline{\mathrm{MS}}}
\def\TR{{\displaystyle \mathrm{T}_{F}}}
\def\gJJ{\gamma_{s}}
\def\gFJ{\gamma_{{\scriptscriptstyle FJ}}}
\def\gFF{\gamma_{{\scriptscriptstyle F\tilde{F}}}}

\def\forcer{{\fontfamily{qcr}\selectfont
Forcer}}

\title{Renormalization of the axial current operator in dimensional regularization at four-loop in QCD}

\author{Long Chen and Micha\l{} Czakon}
\emailAdd{longchen@physik.rwth-aachen.de, mczakon@physik.rwth-aachen.de}
\affiliation{Institut f\"ur Theoretische Teilchenphysik und Kosmologie, RWTH Aachen University,\\ Sommerfeldstr.~16, 52056 Aachen, Germany}

\preprint{TTK-21-54, P3H-21-099}

\abstract{
We provide the renormalization constants of the axial current operators, both singlet and non-singlet, in dimensional regularization up to four-loop order in  (QCD), determined using the off-shell Ward-Takahashi identity for an axial current with a non-anticommuting $\gamma_5$.
A possible application of the result for the singlet axial current operator is the extraction of the non-decoupling mass logarithms in the axial quark form factors.
}

\begin{document}
\allowdisplaybreaks[4]
\unitlength1cm
\maketitle
\flushbottom

\section{Introduction}
\label{sec:intro}

Dimensional regularization (DR)~\cite{tHooft:1972tcz,Bollini:1972ui} is the regularization framework that underlies most of the modern higher-order perturbative calculations in Quantum Chromodynamics (QCD). It is well known that special attention is needed in DR in the treatment of $\gamma_5$ -- an intrinsically 4-dimensional object.
At the root of the issue is the contradiction between a fully anticommuting (non-trivial) $\gamma_5$ in D ($\neq 4$) dimensions and the non-vanishing value of the trace of the product of four $\gamma$ matrices and $\gamma_5$ in 4 dimensions. Nevertheless, the anticommutativity  of $\gamma_5$ is essential for the concept of chirality of spinors in 4 dimensions.
On the other hand, an anticommuting $\gamma_5$ in DR, where the invariance of loop integrals under arbitrary loop-momentum shifts is ensured, leads to the absence of the axial or Adler-Bell-Jackiw (ABJ) anomaly~\cite{Adler:1969gk,Bell:1969ts}.
In order to overcome these issues, various $\gamma_5$ prescriptions in DR have been developed in the literature~\cite{tHooft:1972tcz,Akyeampong:1973xi,Breitenlohner:1977hr,Bardeen:1972vi,Chanowitz:1979zu,Gottlieb:1979ix,Ovrut:1981ne,Espriu:1982bw,Buras:1989xd,Kreimer:1989ke,Korner:1991sx,Larin:1991tj,Larin:1993tq,Jegerlehner:2000dz,Moch:2015usa,Zerf:2019ynn} over a span of nearly 50 years, albeit each with its own pros and cons.

Given the fact that a local gauge theory with an internal gauge anomaly suffers from severe theoretical problems, e.g.~the loss of unitarity and all-order (multiplicative) renormalizability, it is natural to wonder why one would care about studying the anomalous singlet axial current at all.
Putting aside many related interesting theoretical developments, an external axial anomaly (i.e.~anomaly in the divergence of an external axial current which is not coupled to the quantized gauge bosons in question) is not only allowed in a gauge theory, but also crucial in understanding some important physical observations, such as a low-energy theorem for $\pi^0 \to \gamma \gamma$ decay~\cite{Sutherland:1967vf,Veltman:1967,Adler:1969gk,Bell:1969ts} and the solution of the so-called $U(1)$ problem~\cite{Weinberg:1975ui,tHooft:1976snw,tHooft:1986ooh}.
Besides, there are practical applications of the renormalization of the anomalous singlet axial current. One example is the determination of the structure of the non-decoupling (heavy-quark) mass logarithms in the quark form factors~\cite{Collins:1978wz,Chetyrkin:1993jm,Chetyrkin:1993ug,Larin:1993ju,Larin:1994va,Chen:2021rft}, which are present in an anomaly-free theory such as the Standard Model. 
These non-decoupling mass logarithms can be resummed by solving the corresponding renormalization group equation where the anomalous dimension is precisely given by that of the renormalized anomalous singlet axial current.
Another application concerns the calculation of polarized proton structure functions (or parton distribution functions) and polarized splitting functions~\cite{Matiounine:1998re,Kodaira:1998jn,Vogt:2008yw,Moch:2014sna,Behring:2019tus,Tarasov:2020cwl,Tarasov:2021yll,Blumlein:2021ryt}.

In this publication, we provide the renormalization constants of both the singlet and non-singlet axial currents regularized with a non-anticommuting $\gamma_5$~\cite{tHooft:1972tcz,Akyeampong:1973xi,Breitenlohner:1977hr} up to the four-loop order in QCD.
We employ, in particular, the  non-anticommuting $\gamma_5$ variant as prescribed in refs.~\cite{Larin:1991tj,Larin:1993tq}.
The renormalization constant for a flavor non-singlet axial current was previously determined in DR to $\mathcal{O}(\alpha_s^3)$ in refs.~\cite{Larin:1991tj,Larin:1993tq}, whereas the finite part of the renormalization constant of the flavor-singlet axial current was computed only to $\mathcal{O}(\alpha_s^2)$ due to the limitation of the approach in use (see below). 
The missing $\mathcal{O}(\alpha_s^3)$ finite part was completed in ref.~\cite{Ahmed:2021spj} where a derivative-free projector was composed in order to efficiently project out the non-vanishing ``anomalous'' form factor of the famous vector--vector--axial-vector (VVA)-amplitude~\cite{Adler:1969gk,Adler:1969er}, i.e.~the matrix element of the axial current operator between the vacuum and a pair of external gluons, at zero-momentum  (where the axial anomaly vanishes).
The choice of the VVA-amplitude is advantageous for determining the $\overline{\mathrm{MS}}$ renormalization constant associated with the axial-anomaly operator (and subsequently verifying its equality with that of the strong coupling constant $\alpha_s$). This choice is, however, not very economical for studying axial current renormalization.

The approach of refs.~\cite{Gorishnii:1985xm,Larin:1991tj,Larin:1993tq} can be used straightforwardly to obtain the renormalization constant for a non-singlet axial current at $\mathcal{O}(\alpha_s^4)$, provided the propagator-type four-loop integrals can be efficiently evaluated which is indeed the case thanks to \forcer~\cite{Ruijl:2017cxj} and the analytic results for the values of the relevant master integrals given in refs.~\cite{Baikov:2010hf,Lee:2011jt}.
However, this approach does not allow to obtain the renormalization constant for a singlet axial current at $\mathcal{O}(\alpha_s^4)$ via a four-loop calculation, because the axial anomaly vanishes at zero momentum in the matrix element of the axial current operator between the vacuum and a pair of external quarks.
Alternatively, in order to retain a non-vanishing anomaly in this matrix element, one has to allow a non-zero momentum flowing through the singlet axial current.
In consequence, in order to deal with a propagator-type kinematic configuration (and to avoid dealing with infrared/collinear divergences), one of the external quarks must be off-shell, carrying away the same amount of input momentum, while the other carries zero momentum.
This naturally leads us to propose to use the Ward-Takahashi identity for an axial current with off-shell quark fields to determine the renormalization constant for a flavor singlet (and non-singlet) axial current at $\mathcal{O}(\alpha_s^4)$ via a four-loop calculation.
In this approach, the non-anticommuting $\gamma_5$ matrix in the contact terms (due to off-shell external quark fields) does not require any renormalization.
We will address this in detail later in the article.
~\\

The article is organized as follows.
In the next section, we recapitulate the basic renormalization properties of the local composite operators involved in our calculation, mainly to lay down the notations and conventions in use.
Section~\ref{sec:WTI} is devoted to a detailed account of the core of our recipe, the usage of the off-shell Ward-Takahashi identity for an axial current with a non-anticommuting $\gamma_5$, for efficiently determining the renormalization constants of axial currents, especially for the anomalous singlet one.
The perturbative results for the renormalization constants of singlet and non-singlet axial currents determined in this way at four-loop order in QCD are presented in section~\ref{sec:res}. 
We conclude in section~\ref{sec:conc}.

\section{Preliminaries}
\label{sec:prel}

In this section, we specify the notations and conventions for the quantities considered in the rest of this article. 
We use the non-anticommuting definition of $\gamma_5$ in dimensional regularization, originally introduced by 't Hooft-Veltman~\cite{tHooft:1972tcz} and Breitenlohner-Maison~\cite{Breitenlohner:1977hr}
\begin{align}
\label{eq:gamma5}
	\gamma_5=-\frac{i}{4!}\epsilon^{\mu\nu\rho\sigma}\gamma_{\mu}\gamma_{\nu}\gamma_{\rho}\gamma_{\sigma}\,,
\end{align}
but with the Levi-Civita tensor\footnote{We use the convention $\epsilon^{0123} = -\epsilon_{0123} = +1$.} $\epsilon^{\mu\nu\rho\sigma}$ treated according to refs.~\cite{Larin:1991tj,Zijlstra:1992kj,Larin:1993tq}, sometimes called Larin’s prescription in the literature.
As summarized in refs.~\cite{Larin:1991tj,Larin:1993tq}, the properly renormalized non-singlet and singlet axial current operator with a non-anticommuting $\gamma_5$ in QCD with $n_f$ massless quarks can be written as 
\begin{eqnarray} 
\label{eq:J5uvr_ns}
\left[ J^{a,\mu}_{5,ns}\right]_{R} &=& Z_{ns} \, \bar{\psi}^{B}  \, \gamma^{\mu}\gamma_5 \, t^a \psi^{B} \nonumber\\
&=& Z^{f}_{ns} \, Z^{ms}_{ns} \,  \bar{\psi}^{B}  \, \frac{-i}{3!} \epsilon^{\mu\nu\rho\sigma} \gamma_{\nu} \gamma_{\rho} \gamma_{\sigma} \, t^a \psi^{B} \,,\\
\label{eq:J5uvr_s}
\left[ J^{\mu}_{5,s}\right]_{R} &=& Z_{s}\, \sum_{q} \, \bar{\psi}^{B}_q  \, \gamma^{\mu}\gamma_5 \, \psi^{B}_q \nonumber\\
&=& Z^{f}_{s} \, Z^{ms}_{s} \, \sum_{q}  \bar{\psi}^{B}_q  \, \frac{-i}{3!} \epsilon^{\mu\nu\rho\sigma} \gamma_{\nu} \gamma_{\rho} \gamma_{\sigma} \, \psi^{B}_q \,,
\end{eqnarray}
where the subscript $R$ at a square bracket denotes operator renormalization.
In the non-singlet current eq.(\ref{eq:J5uvr_ns}), $t^a$ denotes a (normalized) generator of a flavor group in the representation spanned by the bare quark flavor-multiplet $\psi^{B}$.
Each $\psi^{B}_q$ in the singlet current eq.(\ref{eq:J5uvr_s}) denotes a bare quark field of flavor $q$ with mass dimension\footnote{In ref.~\cite{Ahmed:2021spj}, a global factor $\mu^{4-D}$ in the mass scale $\mu$ of dimensional regularization was introduced in order for the mass dimension of the r.h.s.\ operator be equal to the canonical dimension of the l.h.s.\ in four dimensions.}~$(D-1)/2$, and the sum extends over all $n_f$ quark fields active in the theory.
In order to correctly define a hermitian axial current with a non-anticommuting $\gamma_5$, the necessary ``symmetrization'' of the matrix product, $\gamma^{\mu}\gamma_5 \rightarrow \frac{1}{2} \big(\gamma^{\mu}\gamma_5-\gamma_5\gamma^{\mu}\big)$, before substituting eq.~(\ref{eq:gamma5}) is understood in the second equality in eqs.~(\ref{eq:J5uvr_ns},\,\ref{eq:J5uvr_s}).

The non-singlet axial current $J^{a,\mu}_{5,ns}$ is conserved, up to the classically-expected mass term $2m\bar{\psi}i\gamma_5 t^a\psi$ which vanishes for massless quark fields, and hence is non-anomalous.
This property is sufficient to ensure that $J^{a,\mu}_{5,ns}$ should remain un-renormalized if one had used an anticommuting $\gamma_5$, similarly to the non-renormalization of the  vector current $\left[ J^{\mu}\right]_{R}=\bar{\psi}^{B}\gamma^{\mu} \psi^{B}$.
The appearance of the ultraviolet (UV) renormalization constant $Z_{ns} \equiv Z^{f}_{ns} \, Z^{ms}_{ns}$ is thus solely to amend the spurious terms originating from the apparent loss of $\gamma_5$'s anticommutativity in the chosen prescription (\ref{eq:gamma5}), such that the following current conservation equation
\begin{eqnarray} 
\label{eq:NSconservation}
\partial_{\mu}\big[J^{a,\mu}_{5,s} \big]_{R} = 0\,,
\end{eqnarray}
holds for the renormalized non-singlet axial current.
In this way, the $\gamma_5$'s anticommutativity is effectively restored for the non-singlet axial current~\cite{Gorishnii:1985xm,Larin:1991tj,Larin:1993tq}.
This also implies that the renormalization constant $Z_{ns} \equiv Z^{f}_{ns} \, Z^{ms}_{ns}$, as a whole, has a vanishing anomalous dimension, namely $\mu^2\frac{\mathrm{d} \ln Z_{ns}}{\mathrm{d} \mu^2} = 0$.

On the other hand, the singlet axial current $J^{\mu}_{5,s}$ in eq.~(\ref{eq:J5uvr_s}) is not conserved, due to the quantum anomaly~\cite{Adler:1969gk,Bell:1969ts}, and does require genuine UV renormalization regardless of the $\gamma_5$ prescription in use. 
Furthermore, it is known to renormalize multiplicatively~\cite{Adler:1969gk,Trueman:1979en}. 
The factor $Z_{s} \equiv Z^{f}_{s} \, Z^{ms}_{s}$ denotes the corresponding UV renormalization constant,
conveniently parameterized as the product of a pure $\MSbar$-renormalization factor  $Z^{ms}_{s}$ and an additional finite renormalization factor $Z^{f}_{s}$.
The latter is needed to restore the correct form of the axial Ward identity, or rather the all-order axial-anomaly equation~\cite{Adler:1969gk,Adler:1969er}, which in terms of renormalized local composite operators reads 
\begin{eqnarray} 
\label{eq:ABJanomalyEQ}
\big[\partial_{\mu} J^{\mu}_{5,s} \big]_{R} = a_s\, n_f\, \TR \,  \big[F \tilde{F} \big]_{R}\,,
\end{eqnarray}
where $\TR=1/2$, $F \tilde{F} \equiv  - \epsilon^{\mu\nu\rho\sigma} F^a_{\mu\nu} F^a_{\rho\sigma} = \epsilon_{\mu\nu\rho\sigma} F^a_{\mu\nu} F^a_{\rho\sigma}$  denotes the contraction of the field strength tensor $F^a_{\mu\nu} = \partial_{\mu} A_{\nu}^{a} - \partial_{\nu} A_{\mu}^{a} + g_s \,  f^{abc} A_{\mu}^{b} A_{\nu}^{c}$ of the gluon field $A_\mu^a$ with its \textit{dual} form.
We use the shorthand notation $a_s \equiv \frac{\alpha_s}{4 \pi} = \frac{g_s^2}{16 \pi^2}$ for the QCD coupling, and $f^{abc}$ denotes the structure constants of the non-Abelian color group of QCD. 
The renormalization-group (RG) equation of $a_s$ in $D$ dimensions reads 
\begin{eqnarray}
\label{eq:beta}
\mu^2\frac{\mathrm{d} \ln a_s}{\mathrm{d} \mu^2} = -\epsilon - \mu^2\frac{\mathrm{d} \ln Z_{a_s}}{\mathrm{d} \mu^2} \equiv -\epsilon + \beta \,,
\end{eqnarray}
where $Z_{a_s}$ stands for the $\MSbar$ coupling renormalization constant and $\beta \equiv - \mu^2\frac{\mathrm{d} \ln Z_{a_s}}{\mathrm{d} \mu^2}$ denotes the QCD beta function, i.e.~the anomalous dimension of the renormalized $a_s$ in 4 dimensions ($\mu$ denotes the mass scale in dimensional regularization).
In contrast to the l.h.s.\ of eq.~\eqref{eq:ABJanomalyEQ}, the renormalization of the axial-anomaly operator $F \tilde{F}$ is not strictly multiplicative (as known from ref.~\cite{Adler:1969gk}), but involves mixing with the divergence of the axial current operator~\cite{Espriu:1982bw,Breitenlohner:1983pi}, 
\begin{eqnarray} 
\label{eq:FFuvr}
\big[F \tilde{F} \big]_{R} = Z_{F\tilde{F}} \, \big[F \tilde{F} \big]_{B} \,+\, 
 Z_{FJ} \, \big[\partial_{\mu} J^{\mu}_{5,s} \big]_{B} \,,
\end{eqnarray}
where the subscript $B$ implies that the fields in the local composite operators are bare.
In the computation of the matrix elements of the axial-anomaly operator $F \tilde{F}$, we employ its equivalent form in terms of the divergence of the Chern-Simons current $K^{\mu}$, namely 
\begin{eqnarray} 
\label{eq:Kcurrent}
F \tilde{F}  &=& \partial_{\mu} K^{\mu} \nonumber\\
 &=& \partial_{\mu} \left(-4 \,\epsilon^{\mu\nu\rho\sigma} \,\left(A_{\nu}^{a} \partial_{\rho} A_{\sigma}^{a} \,+\, g_s\,\frac{1}{3} f^{abc} A_{\nu}^{a} A_{\rho}^{b} A_{\sigma}^{c} \right) \right)\,,
\end{eqnarray}
where, unlike $J^{\mu}_{5,s}$, the current $K^{\mu} \equiv -4 \, \epsilon^{\mu\nu\rho\sigma} \,\left(A_{\nu}^{a} \partial_{\rho} A_{\sigma}^{a} \,+\, g_s\,\frac{1}{3} f^{abc} A_{\nu}^{a} A_{\rho}^{b} A_{\sigma}^{c} \right)$ is not a gauge-invariant object~\cite{Espriu:1982bw,Breitenlohner:1983pi}.
The Feynman rules for the r.h.s. of eq.~\eqref{eq:ABJanomalyEQ} used in our calculation are directly based on eq.~\eqref{eq:Kcurrent}. 
As verified explicitly to $\mathcal{O}(\alpha_s^4)$ in ref.~\cite{Ahmed:2021spj}, the equality $Z_{F\tilde{F}} = Z_{a_s}$ holds, which was proved recently in ref.~\cite{Luscher:2021bog}.
For the sake of self-explanatory notations, we keep the symbol $Z_{F\tilde{F}}$ in the following discussion.

The renormalization of the operators $\partial_{\mu} J^{\mu}_{5}$ and $F \tilde{F}$ specified above can be arranged into the following matrix form 
\begin{eqnarray}
\label{eq:Zsmatrix}
\begin{pmatrix}
\big[\partial_{\mu} J^{\mu}_{5,s} \big]_{R}\\
\big[F \tilde{F}\big]_{R}
\end{pmatrix}
= 
\begin{pmatrix}
Z_{s} &  0 \\
Z_{FJ} &  Z_{F\tilde{F}}
\end{pmatrix}
\cdot 
\begin{pmatrix}
\big[\partial_{\mu} J^{\mu}_{5,s} \big]_{B}\\
\big[F \tilde{F}\big]_{B}
\end{pmatrix}\,.
\end{eqnarray}
The matrix of anomalous dimensions of these two renormalized operators is defined by 
\begin{eqnarray}
\label{eq:AMDmatrix}
\frac{\mathrm{d}}{\mathrm{d}\, \ln \mu^2}\,
\begin{pmatrix}
\big[\partial_{\mu} J^{\mu}_{5,s} \big]_{R}\\
\big[F \tilde{F}\big]_{R}
\end{pmatrix}
= 
\begin{pmatrix}
\gJJ &  0 \\
\gFJ &  \gFF
\end{pmatrix}
\cdot 
\begin{pmatrix}
\big[\partial_{\mu} J^{\mu}_{5,s} \big]_{R}\\
\big[F \tilde{F}\big]_{R}
\end{pmatrix} \, .
\end{eqnarray}

\section{The Axial Ward-Takahashi Identity}
\label{sec:WTI}

\begin{figure}[htbp]
\begin{center}
\includegraphics[scale=0.45]{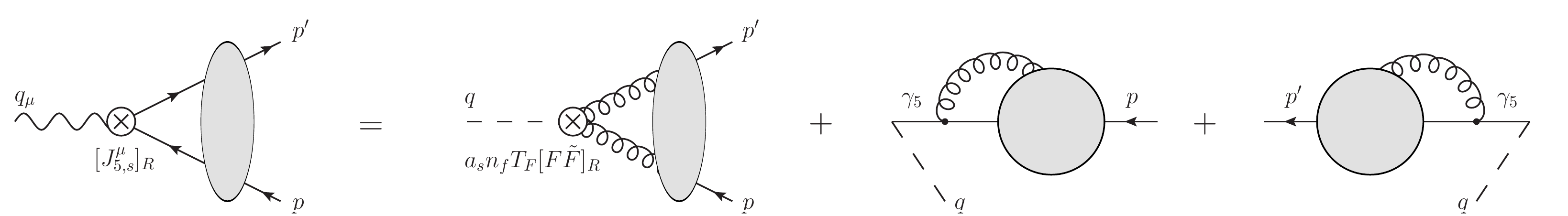}
\caption{An illustration of the anomalous Ward-Takahashi identity (\ref{eq:WTI_s}). The first diagram on the r.h.s.\ is only one of three types of contributions, while the other two involve a local vertex with either three gluons or two quarks as implied by eqs.~\eqref{eq:Kcurrent} and \eqref{eq:Zsmatrix}.
}
\label{fig:AWTI}
\end{center}
\end{figure}

Given the operator-level axial anomaly equation (\ref{eq:ABJanomalyEQ}) in the canonical quantization formalism, one can derive the following anomalous Ward-Takahashi identity~\cite{Adler:1969gk} in the momentum space: %
\begin{eqnarray} 
\label{eq:WTI_s}
q_{\mu}\, \mathrm{\Gamma}^{\mu}_{5,s}(p',p) = -a_s\, n_f\, \TR \,\mathrm{\Lambda}(p',p)   \,+\, \gamma_5\, \hat{S}^{-1}(p) \,+\, \hat{S}^{-1}(p')\, \gamma_5\, ,
\end{eqnarray} %
for the singlet axial current with massless quark fields\footnote{The trivial identity color factor in the fundamental representation is suppressed; in case of massive quark fields, there will be an additional contribution generated by the classically-expected mass term $2m\bar{\psi}i\gamma_5\psi$.}.
Diagrammatically, this equation can be illustrated as in figure~\ref{fig:AWTI}. %
$\mathrm{\Gamma}^{\mu}_{5,s}(p',p)$ in eq.~(\ref{eq:WTI_s}) is the amputated one-particle irreducible (1PI) 3-point vertex function, computed using the renormalized singlet axial current operator (i.e.~eq.~(\ref{eq:J5uvr_s})) with a momentum insertion $q = p' - p$.
$\hat{S}(p)$ in the contact terms on the r.h.s. is defined by the full propagator of an off-shell massless quark, %
\begin{center}
$i \,\delta_{ij}\, \hat{S}(p) = i \,\delta_{ij}\,\frac{\, Z_2}{\slashed{p} - \mathrm{\Sigma}_R(p)}$     \, , 
\end{center} %
where $Z_2$ stands for the quark wavefunction renormalization constant,  $\mathrm{\Sigma}_R(p)$ is the renormalized quark self-energy function and $\delta_{ij}$ is the identity color factor in the fundamental representation of the color algebra.
$\mathrm{\Lambda}(p',p)$ denotes the amputated 1PI 3-point function 
\begin{center} %
$\int d^4 x  d^4 x' \, e^{i p' \cdot x' - i p \cdot x }\,  \langle 0| \hat{\mathrm{T}}\left[ \psi_B(x') \, \bar{\psi}_B(x) \, \big[F \tilde{F} \big]_{R}(0) \right] |0 \rangle |_{\mathrm{amp}}$  \, ,
\end{center} %
computed using the renormalized axial anomaly operator, i.e.~the r.h.s. of eq.~(\ref{eq:ABJanomalyEQ}).

The non-anomalous Ward-Takahashi identity for the renormalized non-singlet axial current operator takes the usual form where the anomaly term in eq.~(\ref{eq:WTI_s}) is absent, namely, 
\begin{eqnarray} 
\label{eq:WTI_ns}
q_{\mu}\, \mathrm{\Gamma}^{a,\mu}_{5,ns}(p',p) = t^a \gamma_5\, \hat{S}^{-1}(p) \,+\, \hat{S}^{-1}(p')\, \gamma_5 t^a \, ,
\end{eqnarray}
which is very similar to the well-known Ward-Takahashi identity for a (color-neutral) conserved vector current.
Similarly, $\mathrm{\Gamma}^{a,\mu}_{5,ns}(p',p)$ in eq.~(\ref{eq:WTI_ns}) is the amputated 1PI vertex function computed using the renormalized non-singlet axial current operator (i.e.~eq.~(\ref{eq:J5uvr_ns})) with a momentum insertion $q = p' - p$.
~\\

One critical feature of eq.(\ref{eq:WTI_s}) and eq.(\ref{eq:WTI_ns}), important for our approach, deserves a comment: 
the $\gamma_5$ matrix in the r.h.s.\ contact terms multiplies the quark self-energy functions from \textit{outside}, hence never appears within any loop correction, and clearly never on a closed fermion loop as illustrated in figure~\ref{fig:AWTI}.
Indeed, the $\gamma_5$ matrix in the contact terms of the axial Ward-Takahashi identities~(\ref{eq:J5uvr_ns},\,\ref{eq:J5uvr_s}) is always on an open quark line and essentially ``external''. 
Because of this, one can simply shift it, while respecting anticommutativity, into the external projector used in the computation, and subsequently combine it with the $\gamma_5$ present in the said external projector.
Therefore, one will not need any $\gamma_5$-related renormalization constant, such as those in eqs.~(\ref{eq:J5uvr_ns},\,\ref{eq:J5uvr_s}), in the calculation of the final renormalized contributions from the contact terms in the axial Ward-Takahashi identities (\ref{eq:WTI_s},\,\ref{eq:WTI_ns}).
In particular, if one would combine these two $\gamma_5$ matrices using the original definition by 't Hooft-Veltman~\cite{tHooft:1972tcz} and Breitenlohner-Maison~\cite{Breitenlohner:1977hr}, one would have $\gamma_5^2$ equal to an identity matrix in spinor space, and hence each of the r.h.s.\ contact terms would contribute the same value as the self-energy function of an off-shell quark without any $\gamma_5$ matrices involved.
Alternatively, if one proceeds literally according to the variant in refs.~\cite{Larin:1991tj,Larin:1993tq}, $\gamma_5^2$ would no longer be the identity matrix, albeit, still proportional to it by a polynomial factor in $D$.
Consequently, the intermediate singular expressions on the r.h.s. would not be identical to the self-energy function of an off-shell quark; but the same finite remainder in the 4-dimensional limit will be produced (without the need of any $\gamma_5$-related renormalization constant).
~\\

To determine the renormalization constants defined in eqs.~(\ref{eq:J5uvr_ns},\,\ref{eq:J5uvr_s}), we compute first the bare expressions of both sides of the axial Ward-Takahashi identities eq.(\ref{eq:WTI_s}) and eq.(\ref{eq:WTI_ns}).
The perturbative QCD corrections to these matrix elements or correlation functions are computed in terms of Feynman diagrams, which are manipulated in the usual way.
The technical workflow follows closely our previous work~\cite{Ahmed:2021spj} where we refer for details, in particular the tool chain in use.
As mentioned in the introduction, to have a non-vanishing divergence of the axial current, namely l.h.s of eqs.~(\ref{eq:J5uvr_ns},\,\ref{eq:J5uvr_s}), one has to allow a non-zero momentum insertion through the axial current operator. 
On the other hand, to keep a propagator-type kinematic configuration, and also to avoid dealing with infrared/collinear divergences, one of the external quarks must be off-shell, carrying away the same amount of momentum inserted through the axial current operator. 
This then forces us to take the kinematic configuration $q = -p$ with $p'=0$.
Consequently, the quark self-energy term $\hat{S}^{-1}(p')$ in the r.h.s. of eq.~(\ref{eq:WTI_s}) and eq.~(\ref{eq:WTI_ns}) vanishes completely in dimensional regularization. 
At this kinematic limit, there is essentially just one Lorentz structure $\slashed{q} \gamma_5$ (at least in 4 dimensions).
We project the aforementioned bare correlation functions of both sides onto this common Lorentz structure, obtaining respectively a Lorentz scalar quantity.
The pure $\MSbar$ parts of the renormalization constants in eqs.~(\ref{eq:J5uvr_ns},\,\ref{eq:J5uvr_s}), namely $Z_{s}^{ms}$ and $Z_{ns}^{ms}$, can be determined from the remaining pole parts in these scalar projections after removing those that can be accounted for by the wavefunction renormalization of the external quark fields, by the $\alpha_s$ renormalization and by the renormalization of the general-covariant-gauge fixing parameter $\xi$.
The non-$\MSbar$ finite renormalization constants, $Z_{s}^{f}$ and $Z_{ns}^{f}$, are determined by demanding the renormalized axial Ward-Takahashi identities at $q = -p$ to hold order-by-order in the perturbative coupling $a_s$ in the 4 dimensional limit.

\section{Results}
\label{sec:res}

Below we present our final perturbative results for the renormalization constants defined in eqs.~(\ref{eq:J5uvr_ns},\,\ref{eq:J5uvr_s}) for axial currents, as well as the $Z_{FJ}$ in eq.(\ref{eq:Zsmatrix}), at the four-loop order in QCD with $n_f$ massless quarks. 
The perturbative QCD coupling $a_s \equiv \frac{\alpha_s}{4 \pi}$ is renormalized in the $\MSbar$ scheme.

\subsection{The non-singlet axial current}
\label{sec:res_ns}

We start with the simple case, the renormalization constant $Z_{ns} \equiv Z^{f}_{ns} \, Z^{ms}_{ns}$ for the non-anomalous non-singlet axial current.
The result for the $\MSbar$ part reads: %
\begin{eqnarray}
\label{eq:Zms_ns}
Z^{ms}_{ns} &=& 1 \,+\, a_s^2 \,\Big\{\,C_A C_F \, \left(\frac{22}{3 \epsilon }\right)\,+\,C_F n_f \, \left(-\frac{4}{3 \epsilon }\right)\Big\} 
\nonumber\\&+& 
a_s^3 \, 
\Big\{\,C_A^2 C_F \, \left(\frac{3578}{81 \epsilon }-\frac{484}{27 \epsilon ^2}\right) \,+\, C_A C_F^2 \, \left(-\frac{308}{9 \epsilon }\right) \,+\, C_A C_F n_f \, \left(\frac{176}{27 \epsilon ^2}-\frac{832}{81 \epsilon }\right)
\nonumber\\&+&
C_F^2 n_f \, \left(\frac{32}{9 \epsilon }\right)\,+\,C_F n_f^2 \, \left(\frac{8}{81 \epsilon }-\frac{16}{27 \epsilon ^2}\right)\Big\}
\nonumber\\&+&
a_s^4 \,
\Big\{\,C_A^3 C_F \, \left(\frac{1331}{27 \epsilon ^3}-\frac{26411}{162 \epsilon ^2}+\frac{\frac{36607}{108}-154 \zeta _3}{\epsilon }\right) \,+\,
C_A^2 C_F^2 \, \left(\frac{121}{\epsilon ^2}+\frac{440 \zeta _3-\frac{29309}{54}}{\epsilon }\right)
\nonumber\\&+& 
C_A^2 C_F n_f \, \left(-\frac{242}{9 \epsilon ^3}+\frac{631}{9 \epsilon ^2}+\frac{-\frac{4 \zeta _3}{3}-\frac{8029}{81}}{\epsilon }\right)
\,+\,
C_A C_F^3 \, \left(\frac{\frac{935}{6}-264 \zeta _3}{\epsilon }\right)
\nonumber\\&+& 
C_A C_F^2 n_f \, \left(\frac{\frac{2381}{27}-\frac{152 \zeta _3}{3}}{\epsilon }-\frac{88}{3 \epsilon ^2}\right)
\,+\,
C_A C_F n_f^2 \, \left(\frac{44}{9 \epsilon ^3}-\frac{206}{27 \epsilon ^2}+\frac{\frac{16 \zeta _3}{3}+\frac{367}{162}}{\epsilon }\right)
\nonumber\\&+&
C_F^3 n_f \, \left(\frac{48 \zeta _3-\frac{40}{3}}{\epsilon }\right)
\,+\,
C_F^2 n_f^2 \, \left(\frac{4}{3 \epsilon ^2}+\frac{\frac{20}{27}-\frac{16 \zeta _3}{3}}{\epsilon }\right)
\nonumber\\&+&
C_F n_f^3 \, \left(-\frac{8}{27 \epsilon ^3}+\frac{4}{81 \epsilon ^2}+\frac{26}{81 \epsilon }\right)
\Big\}\,. 
\end{eqnarray} %
The definition of the quadratic Casimir color constants is as usual: $C_A = N_c \,, \, C_F = (N_c^2 - 1)/(2 N_c) \,$ with $N_c =3$ in QCD and the color-trace normalization factor $\TR = {1}/{2}$.
The result for the finite part reads: %
\begin{eqnarray}
\label{eq:Zf_ns}
Z^{f}_{ns} &=& 1 \,+\, a_s \, \Big\{\,C_F \, \left(-4\right)\Big\} \,+\, a_s^2 \, \Big\{C_A C_F \, \left(-\frac{107}{9}\right)\,+\,C_F^2 \, \left(22\right)\,+\,C_F n_f \, \left(\frac{2}{9}\right)\Big\}
\nonumber\\&+& 
a_s^3 \,
\Big\{\,C_A^2 C_F \, \left(56 \zeta _3-\frac{2147}{27}\right)\,+\,C_A C_F^2 \, \left(\frac{5834}{27}-160 \zeta _3\right)\,+\,C_A C_F n_f \, \left(\frac{32 \zeta _3}{3}+\frac{356}{81}\right)
\nonumber\\&+& 
C_F^3 \, \left(96 \zeta _3-\frac{370}{3}\right)\,+\,C_F^2 n_f \, \left(-\frac{32 \zeta _3}{3}-\frac{62}{27}\right)\,+\,C_F n_f^2 \, \left(\frac{52}{81}\right)\Big\}
\nonumber\\&+& 
a_s^4 \,
\Big\{\,C_A^3 C_F \, \left(\frac{10498 \zeta _3}{27}-\frac{4120 \zeta _5}{3}-\frac{77 \pi ^4}{30}-\frac{324575}{648}\right)
\nonumber\\&+& 
C_A^2 C_F^2 \, \left(-1570 \zeta _3+\frac{17020 \zeta _5}{3}+\frac{22 \pi ^4}{3}+\frac{10619}{6}\right)
\nonumber\\&+& 
C_A^2 C_F n_f \, \left(\frac{1118 \zeta _3}{9}-\frac{200 \zeta _5}{3}-\frac{\pi ^4}{45}+\frac{18841}{324}\right)
\nonumber\\&+&
C_A C_F^3 \, \left(\frac{3332 \zeta _3}{3}-6760 \zeta _5-\frac{22 \pi ^4}{5}-\frac{232949}{108}\right)
\nonumber\\&+&
C_A C_F^2 n_f \, \left(-\frac{1312 \zeta _3}{9}-\frac{40 \zeta _5}{3}-\frac{38 \pi ^4}{45}-\frac{1705}{81}\right)
\nonumber\\&+&
C_A C_F n_f^2 \, \left(-\frac{88 \zeta _3}{9}+\frac{4 \pi ^4}{45}+\frac{73}{54}\right)
\,+\,
C_F^4 \, \left(564 \zeta _3+1840 \zeta _5+\frac{1553}{2}\right)
\nonumber\\&+&
C_F^3 n_f \, \left(\frac{116 \zeta _3}{3}+80 \zeta _5+\frac{4 \pi ^4}{5}-\frac{850}{27}\right)
\,+\,
C_F^2 n_f^2 \, \left(\frac{88 \zeta _3}{9}-\frac{4 \pi ^4}{45}-\frac{719}{81}\right)
\nonumber\\&+&
C_F n_f^3 \, \left(\frac{5}{9}-\frac{16 \zeta _3}{27}\right)
\,+\,
C_1 C_F n_f \, \left(128 \zeta _3-\frac{608}{3}\right)
\nonumber\\&+&
C_2 C_F \, \left(1520 \zeta _3+1920 \zeta _5-\frac{32}{3}\right)\Big\}
\,,
\end{eqnarray} %
where the additional color constants are defined in terms of symmetric color tensors\footnote{The symmetric tensor $d_F^{abcd}$ is defined by the color trace $\frac{1}{6} \mbox{Tr}$~$\big( T^a T^b T^c T^d + T^a T^b T^d T^c + T^a T^c T^b T^d + T^a T^c T^d T^b + T^a T^d T^b T^c + T^a T^d T^c T^b \big)$ with $T^a$ the generators of the fundamental representation of the SU($N_c$) group, and similarly $d_A^{abcd}$ for the adjoint representation.} as %
\begin{eqnarray}
C_1 &\equiv& \frac{d_F^{abcd} d_F^{abcd}}{N_c^2-1} = \frac{N_c^4-6 N_c^2+18}{96 N_c^2} 
\,,\,\nonumber\\
C_2 &\equiv& \frac{d_F^{abcd} d_A^{abcd}}{N_c^2-1} = \frac{N_c \left(N_c^2+6\right)}{48}\,.
\end{eqnarray}
Up to three-loop order, we have reproduced the results in ref.~\cite{Larin:1991tj}.
We have checked that the anomalous dimension of the above $Z_{ns}$ vanishes up to four-loop order (in the 4-dimensional limit), as expected from eq.~(\ref{eq:NSconservation}).
In addition, we performed an alternative calculation of $Z_{ns}$ following the recipe of refs.~\cite{Gorishnii:1985xm,Larin:1991tj,Larin:1993tq} at four-loop order, and we confirm that the same result follows.

\subsection{The singlet axial current}
\label{sec:res_s}

Now we come to the more interesting and useful case of the anomalous singlet axial current operator.
The result for the $\MSbar$ part reads: %
\begin{eqnarray}
\label{eq:Zms_s}
Z^{ms}_{s} &=& 1 \,+\, a_s^2 \,\Big\{\,C_A C_F \, \left(\frac{22}{3 \epsilon }\right)\,+\,C_F n_f \, \left(\frac{5}{3 \epsilon }\right)\Big\}
\nonumber\\&+& 
a_s^3 \, 
\Big\{\,C_A^2 C_F \, \left(\frac{3578}{81 \epsilon }-\frac{484}{27 \epsilon ^2}\right)\,+\,C_A C_F^2 \, \left(-\frac{308}{9 \epsilon }\right)\,+\,C_A C_F n_f \, \left(\frac{149}{81 \epsilon }-\frac{22}{27 \epsilon ^2}\right)
\nonumber\\&+& C_F^2 n_f \, \left(-\frac{22}{9 \epsilon }\right)\,+\,C_F n_f^2 \, \left(\frac{20}{27 \epsilon ^2}+\frac{26}{81 \epsilon }\right)\Big\}
\nonumber\\&+& 
a_s^4 \,
\Big\{C_A^3 C_F \, \left(\frac{1331}{27 \epsilon ^3}-\frac{26411}{162 \epsilon ^2}+\frac{\frac{36607}{108}-154 \zeta _3}{\epsilon }\right)
\nonumber\\&+& 
C_A^2 C_F^2 \, \left(\frac{121}{\epsilon ^2}+\frac{440 \zeta _3-\frac{29309}{54}}{\epsilon }\right)
\,+\,
C_A^2 C_F n_f \, \left(-\frac{121}{18 \epsilon ^3}+\frac{713}{36 \epsilon ^2}+\frac{-\frac{437 \zeta _3}{6}-\frac{15593}{648}}{\epsilon }\right)
\nonumber\\&+& 
C_A C_F^3 \, \left(\frac{\frac{935}{6}-264 \zeta _3}{\epsilon }\right)
\,+\,
C_A C_F^2 n_f \, \left(\frac{55}{6 \epsilon ^2}+\frac{\frac{46 \zeta _3}{3}-\frac{1897}{108}}{\epsilon }\right)
\nonumber\\&+& 
C_A C_F n_f^2 \, \left(-\frac{22}{9 \epsilon ^3}+\frac{76}{27 \epsilon ^2}+\frac{-\frac{53 \zeta _3}{3}-\frac{31}{81}}{\epsilon }\right)
\,+\,
C_F^3 n_f \, \left(\frac{48 \zeta _3+\frac{29}{12}}{\epsilon }\right)
\nonumber\\&+&
C_F^2 n_f^2 \, \left(\frac{11}{6 \epsilon ^2}+\frac{\frac{56 \zeta _3}{3}-\frac{851}{54}}{\epsilon }\right)
\,+\, 
C_F n_f^3 \, \left(\frac{10}{27 \epsilon ^3}+\frac{13}{81 \epsilon ^2}-\frac{35}{162 \epsilon }\right)\Big\}\,.
\end{eqnarray} %
The result for the finite part reads: %
\begin{eqnarray}
\label{eq:Zf_s}
Z^{f}_{s} &=& 1 \,+\, a_s \, \Big\{\,C_F \, \left(-4\right)\Big\} \,+\, a_s^2 \, \Big\{\,C_A C_F \, \left(-\frac{107}{9}\right),C_F^2 \, \left(22\right)\,+\,C_F n_f \, \left(\frac{31}{18}\right)\Big\}
\nonumber\\&+& 
a_s^3 \, 
\Big\{\,C_A^2 C_F \, \left(56 \zeta _3-\frac{2147}{27}\right)\,+\,C_A C_F^2 \, \left(\frac{5834}{27}-160 \zeta _3\right)
\,+\,C_A C_F n_f \, \left(\frac{110 \zeta _3}{3}-\frac{133}{81}\right)
\nonumber\\&+& 
C_F^3 \, \left(96 \zeta _3-\frac{370}{3}\right)
\,+\, C_F^2 n_f \, \left(\frac{497}{54}-\frac{104 \zeta _3}{3}\right)\,+\,C_F n_f^2 \, \left(\frac{316}{81}\right)\Big\}
\nonumber\\&+& 
a_s^4 \,
\Big\{\,C_A^3 C_F \, \left(\frac{10498 \zeta _3}{27}-\frac{4120 \zeta _5}{3}-\frac{77 \pi ^4}{30}-\frac{324575}{648}\right)
\nonumber\\&+& 
C_A^2 C_F^2 \, \left(-1570 \zeta _3+\frac{17020 \zeta _5}{3}+\frac{22 \pi ^4}{3}+\frac{10619}{6}\right)
\nonumber\\&+& 
C_A^2 C_F n_f \, \left(\frac{3941 \zeta _3}{9}-\frac{845 \zeta _5}{4}-\frac{437 \pi ^4}{360}+\frac{98815}{1296}\right)
\nonumber\\&+& 
C_A C_F^3 \, \left(\frac{3332 \zeta _3}{3}-6760 \zeta _5-\frac{22 \pi ^4}{5}-\frac{232949}{108}\right)
\nonumber\\&+& 
C_A C_F^2 n_f \, \left(-\frac{8807 \zeta _3}{18}-\frac{370 \zeta _5}{3}+\frac{23 \pi ^4}{90}+\frac{10687}{162}\right)
\nonumber\\&+& 
C_A C_F n_f^2 \, \left(-\frac{1037 \zeta _3}{18}+\frac{91 \pi ^4}{180}-\frac{235}{36}\right)
\,+\,
C_F^4 \, \left(564 \zeta _3+1840 \zeta _5+\frac{1553}{2}\right)
\nonumber\\&+& 
C_F^3 n_f \, \left(\frac{443 \zeta _3}{3}+260 \zeta _5+\frac{4 \pi ^4}{5}-\frac{43025}{216}\right)
\,+\,
C_F^2 n_f^2 \, \left(\frac{484 \zeta _3}{9}-\frac{22 \pi ^4}{45}-\frac{5971}{648}\right)
\nonumber\\&+& 
C_F n_f^3 \, \left(\frac{17}{4}-\frac{124 \zeta _3}{27}\right)
\,+\, C_1 C_F n_f \, \left(264 \zeta _3-520 \zeta _5-\frac{320}{3}\right)
\nonumber\\&+& 
C_2 C_F \, \left(1520 \zeta _3+1920 \zeta _5-\frac{32}{3}\right)\Big\}\,.
\end{eqnarray} %
Up to three-loop order, we have reproduced the results in ref.~\cite{Larin:1993tq,Ahmed:2021spj} determined by computing the VVA-amplitudes.
As can be easily checked using these results, the differences $Z^{ms}_{s} -Z^{ms}_{ns}$ and $Z^{f}_{s} -Z^{f}_{ns}$ all start from $\mathcal{O}(a_s^2)$ and are proportional to $n_f\, C_F$.
As discussed in section~\ref{sec:prel}, the renormalized singlet axial current does have a non-zero anomalous dimension, which reads: %
\begin{eqnarray}
\label{eq:AMD_Zs}
\gJJ &=&  \mu^2 \frac{\mathrm{d}\, \ln \big( Z^{f}_{s} \, Z^{ms}_{s}\big)}{\mathrm{d}\, \mu^2} \nonumber\\
&=&
a_s \, \Big\{\,C_F \, \left(4 \epsilon\right)\Big\} \,+\,
a_s^2 \, \Big\{C_A C_F \, \left(\frac{214 \epsilon }{9}\right)\,+\,C_F^2 \, \left(-28 \epsilon\right)\,+\,C_F n_f \, \left(-\frac{31 \epsilon }{9}-6\right)\Big\}
\nonumber\\&+& 
a_s^3 \, 
\Big\{\,C_A^2 C_F \, \left(\left(\frac{2147}{9}-168 \zeta _3\right) \epsilon\right)\,+\,C_A C_F^2 \, \left(\left(480 \zeta _3-\frac{4550}{9}\right) \epsilon\right)
\nonumber\\&+& 
C_A C_F n_f \, \left(\left(\frac{133}{27}-110 \zeta _3\right) \epsilon -\frac{142}{3}\right)\,+\,C_F^3 \, \left(\left(170-288 \zeta _3\right) \epsilon\right)
\nonumber\\&+& 
C_F^2 n_f \, \left(\left(104 \zeta _3-\frac{869}{18}\right) \epsilon +18\right)\,+\,C_F n_f^2 \, \left(\frac{4}{3}-\frac{316 \epsilon }{27}\right)\Big\}
\nonumber\\&+& 
a_s^4 \, 
\Big\{\,C_A^3 C_F \, \left(\left(-\frac{41992 \zeta _3}{27}+\frac{16480 \zeta _5}{3}+\frac{154 \pi ^4}{15}+\frac{324575}{162}\right) \epsilon\right)
\nonumber\\&+&
C_A^2 C_F^2 \, \left(\left(5384 \zeta _3-\frac{68080 \zeta _5}{3}-\frac{88 \pi ^4}{3}-\frac{447472}{81}\right) \epsilon\right)
\nonumber\\&+&
C_A^2 C_F n_f \, \left(\left(-\frac{15764 \zeta _3}{9}+845 \zeta _5+\frac{437 \pi ^4}{90}-\frac{98815}{324}\right) \epsilon -\frac{1607}{6}\right)
\nonumber\\&+&
C_A C_F^3 \, \left(\left(-\frac{5648 \zeta _3}{3}+27040 \zeta _5+\frac{88 \pi ^4}{5}+\frac{43967}{9}\right) \epsilon\right)
\nonumber\\&+&
C_A C_F^2 n_f \, \left(\left(\frac{12334 \zeta _3}{9}+\frac{1480 \zeta _5}{3}-\frac{46 \pi ^4}{45}-\frac{25880}{81}\right) \epsilon +\frac{461}{2}\right)
\nonumber\\&+&
C_A C_F n_f^2 \, \left(144 \zeta _3+\left(\frac{2074 \zeta _3}{9}-\frac{91 \pi ^4}{45}+\frac{235}{9}\right) \epsilon -\frac{82}{3}\right)
\nonumber\\&+&
C_F^3 n_f \, \left(\left(-36 \zeta _3-1040 \zeta _5-\frac{16 \pi ^4}{5}+\frac{4145}{6}\right) \epsilon -63\right)
\nonumber\\&+&
C_F^2 n_f^2 \, \left(-144 \zeta _3+\left(-\frac{1936 \zeta _3}{9}+\frac{88 \pi ^4}{45}-\frac{530}{27}\right) \epsilon +107\right)
\nonumber\\&+&
C_F^4 \, \left(\left(-3792 \zeta _3-7360 \zeta _5-\frac{3950}{3}\right) \epsilon\right)
\,+\,
C_F n_f^3 \, \left(\left(\frac{496 \zeta _3}{27}-17\right) \epsilon +\frac{26}{3}\right)
\nonumber\\&+&
C_1 C_F n_f \, \left(\left(-1056 \zeta _3+2080 \zeta _5+\frac{1280}{3}\right) \epsilon\right)
\,+\,
C_2 C_F \, \left(\left(-6080 \zeta _3-7680 \zeta _5+\frac{128}{3}\right) \epsilon\right)\Big\}\,.\nonumber\\
\end{eqnarray} %
We note that due to the appearance of the finite renormalization $Z_s^f$, the anomalous dimension of $Z_s$ given above contains terms linear in $\epsilon$, due to the usage of the D-dimensional QCD beta function defined in eq.~(\ref{eq:beta}).

As mentioned in the introduction, among the practical applications of the knowledge of the renormalization of the anomalous singlet axial current is the determination of the structure of the non-decoupling heavy-quark mass logarithms in the axial quark form factors, which are present in an anomaly-free theory such as the Standard Model. 
It is known~\cite{Collins:1978wz,Chetyrkin:1993jm,Chetyrkin:1993ug,Larin:1993ju,Larin:1994va} that the effect of top-quark loops in axial quark form factors does not decouple in the large top-quark mass or low-energy limit due to the presence of the axial-anomaly type diagrams.
A Wilson coefficient in front of the renormalized singlet axial current operator can be introduced in the low-energy or heavy-top effective Lagrangian, which encodes all non-decoupling top-quark mass logarithms remaining in the total ``physical'' (non-anomalous) quark form factors (See, for instance,  refs.~\cite{Chetyrkin:1993ug,Chen:2021rft} for a detailed discussion).
The terms in this Wilson coefficient featuring the mass logarithms at the four-loop order can be determined from the anomalous dimension of the singlet axial current given in eq.~(\ref{eq:AMD_Zs}) in 4 dimensions, by solving the corresponding RG equation (e.g.~in the form given by eq.~(6.3) of ref.~\cite{Chen:2021rft}), provided the full knowledge of the lower-order results available in refs.~\cite{Chen:2021rft,Ju:2021lah}.
Following the conventions of ref.~\cite{Chen:2021rft}, the renormalized low-energy effective Lagrangian with the axial current coupling to the Z-boson field $\text{Z}_{\mu}$ reads: %
\begin{eqnarray}
\label{eq:Leff}
\delta \mathcal{L}^{R}_{\mathrm{eff}} &=& \mu^\epsilon \Big(
Z_{ns}\,
\sum_{i=1}^{n_f} a_i \,  \bar{\psi}^{B}_{i}  \, \gamma^{\mu}\gamma_5 \, \psi^{B}_i \,+\,
a_b \, Z_{ps} \, \big[ J^{\mu}_{5,s} \big]_B \nonumber\\
&& \,\,\,\,\quad 
+\, a_t\, C_w(a_s, \mu/m_t)\, \big(Z_{ns} + n_f\,Z_{ps} \big) \, \big[ J^{\mu}_{5,s} \big]_B \Big)\, \text{Z}_{\mu} \,,
\end{eqnarray} %
where $C_w(a_s,\mu/m_t)$ with $m_t$ the on-shell top-quark mass is the aforementioned Wilson coefficient. 
Furthermore, the axial electroweak coupling of the quark $i$ is denoted by $a_i$. 
The renormalization constant $Z_{ps} \equiv \frac{1}{n_f} \big(Z_s - Z_{ns} \big)$ is defined as the difference between the non-singlet and singlet axial-current renormalization constants defined respectively in eq.~(\ref{eq:J5uvr_ns}) and eq.~(\ref{eq:J5uvr_s}), further normalized to the case of a single quark flavor with axial coupling.
Here, $n_f = 5$ and $a_s$ is to be taken after decoupling the top quark in the $m_t \rightarrow \infty$ limit.
Following from the RG invariance of the effective Lagrangian eq.~(\ref{eq:Leff}), the RG equation for $C_w(a_s,\mu/m_t)$ reads: %
\begin{eqnarray}
\label{eq:CwRGE}
\mu^2\frac{\mathrm{d}}{\mathrm{d} \mu^2} C_w(a_s, \mu/m_t) 
&=& \mu^2\frac{\partial}{\partial \mu^2} C_w(a_s, \mu/m_t)  \,+\, \beta\, a_s\frac{\partial}{\partial a_s} C_w(a_s, \mu/m_t)  \nonumber\\
&=& \frac{\gamma_s}{n_f} - \gamma_s \,C_w(a_s, \mu/m_t)\,,
\end{eqnarray} 
where only the 4-dimensional limit of $\gamma_s$ in eq.~(\ref{eq:AMD_Zs}) is needed.
Solving eq.~(\ref{eq:CwRGE}) with a perturbative ansatz for $C_w(a_s,\mu/m_t)$, the logarithmic part of this Wilson coefficient can be explicitly reconstructed up to $ \mathcal{O}(\alpha_s^4)$: %
\begin{eqnarray}
\label{eq:CwRes}
C_w(a_s,\mu/m_t)|_{\text{log}} &=&
a_s^2\,\Big\{C_F \, \left(-6 L_{\mu }\right)\Big\} 
\nonumber\\&+& 
a_s^3\,\Big\{C_A C_F \, \Big(-22 L_{\mu }^2-\frac{76 L_{\mu }}{3}\Big)
\,+\, C_F^2 \, \left(18 L_{\mu }\right)
\,+\,
C_F n_f \, \Big(4 L_{\mu }^2-\frac{8 L_{\mu }}{3}\Big)\Big\}
\nonumber\\&+& 
a_s^4\,\Big\{
C_A^2 C_F \, \Big(-\frac{242 L_{\mu }^3}{3}-\frac{622 L_{\mu }^2}{3}-\frac{10868 L_{\mu }}{9}+924 \zeta _3 L_{\mu }\Big)
\,+\,
C_F^3 \, \left(-63 L_{\mu }\right)
\nonumber\\&+& 
C_F n_f \, \Big(-\frac{328 L_{\mu }}{9}\Big)
\,+\,
C_A C_F^2 \, \Big(-792 \zeta _3 L_{\mu }+99 L_{\mu }^2-50 L_{\mu }\Big)
\nonumber\\&+& 
C_F n_f^2 \, \Big(-\frac{8 L_{\mu }^3}{3}+\frac{8 L_{\mu }^2}{3}-\frac{296 L_{\mu }}{9}\Big)
\,+\,
C_A C_F \, \Big(\frac{1804 L_{\mu }}{9}\Big)
\nonumber\\&+& 
C_A C_F n_f \, \Big(\frac{88 L_{\mu }^3}{3}+\frac{92 L_{\mu }^2}{3}+\frac{3280 L_{\mu }}{9}-24 \zeta _3 L_{\mu }\Big)
\nonumber\\&+& 
C_F^2 n_f \, \Big(164 L_{\mu }-24 L_{\mu }^2\Big)
\Big\}\, + \, \mathcal{O}(\alpha_s^5)\,,
\end{eqnarray}
where $L_{\mu} \equiv \ln \frac{\mu^2}{m^2_t}$.\footnote{We note that the number of the heavy decoupled quarks is set as 1 in eq.~(\ref{eq:CwRes}), and this is the reason why there appear terms with factors like $C_A C_F $, $C_F n_f$ in the $\mathcal{O}(\alpha_s^4)$ result.} 
The constant part of $C_w(a_s,\mu/m_t)$, however, needs to be fixed by matching to an explicit calculation in the heavy-top limit.
We plan to come back to this in the future.

\subsection{The axial anomaly operator}
\label{sec:res_aa}

As argued in the literature, e.g.~refs.~\cite{Breitenlohner:1983pi,Bos:1992nd,Larin:1993tq,Ahmed:2021spj,Luscher:2021bog}, it follows from the ABJ equation with the proven relation $Z_{F\tilde{F}} = Z_{a_s}$, that the equality 
\begin{eqnarray}
\label{eq:AMDsEQ}
\gJJ = n_f \, \TR \, a_s\, \gFJ 
\end{eqnarray}
with $\gJJ$ and $\gFJ$ defined in eq.~(\ref{eq:AMDmatrix}) 
must hold true, albeit only in the limit $\epsilon = 0$.
In principle, in the matrix of renormalization constants in eq.(\ref{eq:Zsmatrix}) for the local composite operators appearing in the ABJ equation, one only needs to compute $Z_s$.
This relation also provides a check of our 4-loop result eq.~(\ref{eq:AMD_Zs}), as $Z_{FJ}$ was already known to three-loop order~\cite{Zoller:2013ixa,Ahmed:2015qpa}.

The operator $\big[F \tilde{F} \big]_{R}$ is only required up to $\mathcal{O}(\alpha_s^3)$ in the calculation of $Z_s$ at four-loop order by means of eq.~(\ref{eq:WTI_s}). Nevertheless, it is possible to obtain $Z_{FJ}$ up to $\mathcal{O}(\alpha_s^4)$ using the same techniques as for the other renormalisation constants. Indeed, we evaluate the matrix element of eq.~\eqref{eq:FFuvr} corresponding to the first contribution on the r.h.s.\ in figure~\ref{fig:AWTI} at $q = -p$ with $p'=0$ up to four-loop order as before. In this way we obtain: %
\begin{eqnarray}
\label{eq:ZFJ}
Z_{FJ} &=&  a_s\, \Big\{C_F \, \left(\frac{12}{\epsilon }\right)\Big\} \nonumber\\&+& 
a_s^2\, \Big\{\,C_A C_F \, \left(\frac{142}{3 \epsilon }-\frac{44}{\epsilon ^2}\right)\,+\,C_F^2 \, \left(-\frac{42}{\epsilon }\right)\,+\,C_F n_f \, \left(\frac{8}{\epsilon ^2}-\frac{4}{3 \epsilon }\right)\Big\}
\nonumber\\&+& 
a_s^3\,
\Big\{\,C_A^2 C_F \, \left(\frac{484}{3 \epsilon ^3}-\frac{2378}{9 \epsilon ^2}+\frac{1607}{9 \epsilon }\right)
\,+\,
C_A C_F^2 \, \left(\frac{550}{3 \epsilon ^2}-\frac{2947}{9 \epsilon }\right)
\nonumber\\&+& 
C_A C_F n_f \, \left(-\frac{176}{3 \epsilon ^3}+\frac{568}{9 \epsilon ^2}+\frac{\frac{164}{9}-96 \zeta _3}{\epsilon }\right)
\,+\,
C_F^3 \, \left(\frac{178}{\epsilon }\right)
\nonumber\\&+& 
C_F^2 n_f \, \left(\frac{96 \zeta _3-\frac{548}{9}}{\epsilon }-\frac{16}{3 \epsilon ^2}\right) \,+\, C_F n_f^2 \, \left(\frac{16}{3 \epsilon ^3}-\frac{8}{9 \epsilon ^2}-\frac{52}{9 \epsilon }\right)\Big\}
\nonumber\\&+& 
a_s^4\,
\Big\{\,C_A^3 C_F \, \left(-\frac{5324}{9 \epsilon ^4}+\frac{36256}{27 \epsilon ^3}-\frac{37111}{27 \epsilon ^2}+\frac{\frac{8540 \zeta _3}{9}-1760 \zeta _5+\frac{110140}{81}}{\epsilon }\right)
\nonumber\\&+& 
C_A^2 C_F^2 \, \left(-\frac{726}{\epsilon ^3}+\frac{50320}{27 \epsilon ^2}+\frac{-800 \zeta _3+1760 \zeta _5-\frac{200645}{81}}{\epsilon }\right)
\nonumber\\&+& 
C_A^2 C_F n_f \, \left(\frac{968}{3 \epsilon ^4}-\frac{5270}{9 \epsilon ^3}+\frac{352 \zeta _3+\frac{3052}{9}}{\epsilon ^2}+\frac{-\frac{5060 \zeta _3}{3}+800 \zeta _5+\frac{88 \pi ^4}{15}-\frac{35143}{162}}{\epsilon }\right)
\nonumber\\&+& 
C_A C_F^2 n_f \, \left(\frac{220}{3 \epsilon ^3}+\frac{\frac{2143}{27}-352 \zeta _3}{\epsilon ^2}+\frac{\frac{5258 \zeta _3}{3}+160 \zeta _5-\frac{88 \pi ^4}{15}-\frac{77101}{162}}{\epsilon }\right)
\nonumber\\&+& 
C_A C_F n_f^2 \, \left(-\frac{176}{3 \epsilon ^4}+\frac{196}{3 \epsilon ^3}+\frac{\frac{191}{9}-64 \zeta _3}{\epsilon ^2}+\frac{\frac{272 \zeta _3}{3}-\frac{16 \pi ^4}{15}+\frac{1954}{81}}{\epsilon }\right)
\nonumber\\&+& 
C_F^4 \, \left(\frac{288 \zeta _3-\frac{1397}{2}}{\epsilon }\right)
\,+\, 
C_F^3 n_f \, \left(\frac{22}{3 \epsilon ^2}+\frac{-\frac{280 \zeta _3}{3}-960 \zeta _5+\frac{5471}{9}}{\epsilon }\right)
\nonumber\\&+& 
C_F^2 n_f^2 \, \left(\frac{32}{3 \epsilon ^3}+\frac{64 \zeta _3-\frac{1352}{27}}{\epsilon ^2}+\frac{-\frac{272 \zeta _3}{3}+\frac{16 \pi ^4}{15}-\frac{1874}{81}}{\epsilon }\right)
\nonumber\\&+& 
C_A C_F^3 \, \left(\frac{\frac{18781}{9}-\frac{1088 \zeta _3}{3}}{\epsilon }-\frac{2728}{3 \epsilon ^2}\right)
\,+\, 
C_F n_f^3 \, \left(\frac{32}{9 \epsilon ^4}-\frac{16}{27 \epsilon ^3}-\frac{104}{27 \epsilon ^2}+\frac{\frac{64 \zeta _3}{9}-\frac{20}{3}}{\epsilon }\right)\Big\}\,.\nonumber\\
\end{eqnarray} %

\section{Conclusion}
\label{sec:conc}

We have computed the renormalization constants of the axial currents, both singlet and non-singlet, in dimensional regularization up to four-loop order in QCD, using the off-shell axial Ward-Takahashi identities with a non-anticommuting $\gamma_5$.
We determined, in addition, the mixing renormalization constant $Z_{FJ}$ at the four-loop order by explicit diagrammatic computation.
With our recipe provided in this article, it seems to even be possible to determine the renormalization constants for a flavor singlet (and non-singlet) axial current at $\mathcal{O}(\alpha_s^5)$ via a five-loop calculation, provided the propagator-type five-loop integrals can be evaluated. 

\section*{Acknowledgements}

The work of L.C. and M.C. was supported by the Deutsche Forschungsgemeinschaft under grant 396021762 -- TRR 257.

\bibliography{Z5} 

\providecommand{\href}[2]{#2}\begingroup\raggedright\begin{thebibliography}{10}

\bibitem{tHooft:1972tcz}
G.~'t~Hooft and M.~J.~G. Veltman, {\it {Regularization and Renormalization of
  Gauge Fields}},
\href{http://dx.doi.org/10.1016/0550-3213(72)90279-9}{{\em Nucl. Phys.}
  {\bfseries B44} (1972) 189--213}.

\bibitem{Bollini:1972ui}
C.~G. Bollini and J.~J. Giambiagi, {\it {Dimensional Renormalization: The
  Number of Dimensions as a Regularizing Parameter}},
\href{http://dx.doi.org/10.1007/BF02895558}{{\em Nuovo Cim.} {\bfseries B12}
  (1972) 20--26}.

\bibitem{Adler:1969gk}
S.~L. Adler, {\it {Axial vector vertex in spinor electrodynamics}},
  \href{http://dx.doi.org/10.1103/PhysRev.177.2426}{{\em Phys. Rev.} {\bfseries
  177} (1969) 2426--2438}.

\bibitem{Bell:1969ts}
J.~S. Bell and R.~Jackiw, {\it {A PCAC puzzle: $\pi^0 \to \gamma \gamma$ in the
  $\sigma$ model}},
\href{http://dx.doi.org/10.1007/BF02823296}{{\em Nuovo Cim.} {\bfseries A60}
  (1969) 47--61}.

\bibitem{Akyeampong:1973xi}
D.~A. Akyeampong and R.~Delbourgo, {\it {Dimensional regularization, abnormal
  amplitudes and anomalies}},
\href{http://dx.doi.org/10.1007/BF02786835}{{\em Nuovo Cim.} {\bfseries A17}
  (1973) 578--586}.

\bibitem{Breitenlohner:1977hr}
P.~Breitenlohner and D.~Maison, {\it {Dimensional Renormalization and the
  Action Principle}},
\href{http://dx.doi.org/10.1007/BF01609069}{{\em Commun. Math. Phys.}
  {\bfseries 52} (1977) 11--38}.

\bibitem{Bardeen:1972vi}
W.~A. Bardeen, R.~Gastmans, and B.~E. Lautrup, {\it {Static quantities in
  Weinberg's model of weak and electromagnetic interactions}},
\href{http://dx.doi.org/10.1016/0550-3213(72)90218-0}{{\em Nucl. Phys.}
  {\bfseries B46} (1972) 319--331}.

\bibitem{Chanowitz:1979zu}
M.~S. Chanowitz, M.~Furman, and I.~Hinchliffe, {\it {The Axial Current in
  Dimensional Regularization}},
\href{http://dx.doi.org/10.1016/0550-3213(79)90333-X}{{\em Nucl. Phys.}
  {\bfseries B159} (1979) 225--243}.

\bibitem{Gottlieb:1979ix}
S.~A. Gottlieb and J.~T. Donohue, {\it {The Axial Vector Current and
  Dimensional Regularization}},
\href{http://dx.doi.org/10.1103/PhysRevD.20.3378}{{\em Phys. Rev.} {\bfseries
  D20} (1979) 3378}.

\bibitem{Ovrut:1981ne}
B.~A. Ovrut, {\it {Axial Vector Ward Identities and Dimensional
  Regularization}},
\href{http://dx.doi.org/10.1016/0550-3213(83)90511-4}{{\em Nucl. Phys.}
  {\bfseries B213} (1983) 241--265}.

\bibitem{Espriu:1982bw}
D.~Espriu and R.~Tarrach, {\it {Renormalization of the Axial Anomaly
  Operators}},
\href{http://dx.doi.org/10.1007/BF01573750}{{\em Z. Phys.} {\bfseries C16}
  (1982) 77}.

\bibitem{Buras:1989xd}
A.~J. Buras and P.~H. Weisz, {\it {QCD Nonleading Corrections to Weak Decays in
  Dimensional Regularization and 't Hooft-Veltman Schemes}},
\href{http://dx.doi.org/10.1016/0550-3213(90)90223-Z}{{\em Nucl. Phys.}
  {\bfseries B333} (1990) 66--99}.

\bibitem{Kreimer:1989ke}
D.~Kreimer, {\it {The $\gamma_5$ Problem and Anomalies: A Clifford Algebra
  Approach}},
\href{http://dx.doi.org/10.1016/0370-2693(90)90461-E}{{\em Phys. Lett.}
  {\bfseries B237} (1990) 59--62}.

\bibitem{Korner:1991sx}
J.~G. K{\"o}rner, D.~Kreimer, and K.~Schilcher, {\it {A Practicable $\gamma_5$
  scheme in dimensional regularization}},
\href{http://dx.doi.org/10.1007/BF01559471}{{\em Z. Phys.} {\bfseries C54}
  (1992) 503--512}.

\bibitem{Larin:1991tj}
S.~A. Larin and J.~A.~M. Vermaseren, {\it {The $\alpha_s^3$ corrections to the
  Bjorken sum rule for polarized electroproduction and to the Gross-Llewellyn
  Smith sum rule}},
\href{http://dx.doi.org/10.1016/0370-2693(91)90839-I}{{\em Phys. Lett.}
  {\bfseries B259} (1991) 345--352}.

\bibitem{Larin:1993tq}
S.~A. Larin, {\it {The Renormalization of the axial anomaly in dimensional
  regularization}},  \href{http://dx.doi.org/10.1016/0370-2693(93)90053-K}{{\em
  Phys. Lett.} {\bfseries B303} (1993) 113--118},
\href{http://arxiv.org/abs/hep-ph/9302240}{{\ttfamily arXiv:hep-ph/9302240
  [hep-ph]}}.

\bibitem{Jegerlehner:2000dz}
F.~Jegerlehner, {\it {Facts of life with $\gamma_5$}},
  \href{http://dx.doi.org/10.1007/s100520100573}{{\em Eur. Phys. J.} {\bfseries
  C18} (2001) 673--679},
\href{http://arxiv.org/abs/hep-th/0005255}{{\ttfamily arXiv:hep-th/0005255
  [hep-th]}}.

\bibitem{Moch:2015usa}
S.~Moch, J.~A.~M. Vermaseren, and A.~Vogt, {\it {On $\gamma_5$ in higher-order
  QCD calculations and the NNLO evolution of the polarized valence
  distribution}},  \href{http://dx.doi.org/10.1016/j.physletb.2015.07.027}{{\em
  Phys. Lett.} {\bfseries B748} (2015) 432--438},
\href{http://arxiv.org/abs/1506.04517}{{\ttfamily arXiv:1506.04517 [hep-ph]}}.

\bibitem{Zerf:2019ynn}
N.~Zerf, {\it {Fermion Traces Without Evanescence}},
  \href{http://dx.doi.org/10.1103/PhysRevD.101.036002}{{\em Phys. Rev.}
  {\bfseries D101} no.~3, (2020) 036002},
\href{http://arxiv.org/abs/1911.06345}{{\ttfamily arXiv:1911.06345 [hep-ph]}}.

\bibitem{Sutherland:1967vf}
D.~G. Sutherland, {\it {Current algebra and some nonstrong mesonic decays}},
  \href{http://dx.doi.org/10.1016/0550-3213(67)90180-0}{{\em Nucl. Phys. B}
  {\bfseries 2} (1967) 433--440}.

\bibitem{Veltman:1967}
M.~Veltman, {\it {Theoretical aspects of high energy neutrino interactions}},
  \href{http://dx.doi.org/https://doi.org/10.1098/rspa.1967.0193}{{\em Proc. R.
  Soc. Lond. A} {\bfseries 301} (1967) 107--112}.

\bibitem{Weinberg:1975ui}
S.~Weinberg, {\it {The U(1) Problem}},
  \href{http://dx.doi.org/10.1103/PhysRevD.11.3583}{{\em Phys. Rev. D}
  {\bfseries 11} (1975) 3583--3593}.

\bibitem{tHooft:1976snw}
G.~'t~Hooft, {\it {Computation of the Quantum Effects Due to a Four-Dimensional
  Pseudoparticle}},  \href{http://dx.doi.org/10.1103/PhysRevD.14.3432}{{\em
  Phys. Rev. D} {\bfseries 14} (1976) 3432--3450}. [Erratum: Phys.Rev.D 18,
  2199 (1978)].

\bibitem{tHooft:1986ooh}
G.~'t~Hooft, {\it {How Instantons Solve the U(1) Problem}},
  \href{http://dx.doi.org/10.1016/0370-1573(86)90117-1}{{\em Phys. Rept.}
  {\bfseries 142} (1986) 357--387}.

\bibitem{Collins:1978wz}
J.~C. Collins, F.~Wilczek, and A.~Zee, {\it {Low-Energy Manifestations of Heavy
  Particles: Application to the Neutral Current}},
  \href{http://dx.doi.org/10.1103/PhysRevD.18.242}{{\em Phys. Rev. D}
  {\bfseries 18} (1978) 242}.

\bibitem{Chetyrkin:1993jm}
K.~G. Chetyrkin and J.~H. Kuhn, {\it {Complete QCD corrections of order
  alpha-s**2 to the Z decay rate}},
  \href{http://dx.doi.org/10.1016/0370-2693(93)90613-M}{{\em Phys. Lett. B}
  {\bfseries 308} (1993) 127--136}.

\bibitem{Chetyrkin:1993ug}
K.~G. Chetyrkin and O.~V. Tarasov, {\it {The alpha-s**3 corrections to the
  effective neutral current and to the Z decay rate in the heavy top quark
  limit}},  \href{http://dx.doi.org/10.1016/0370-2693(94)91538-5}{{\em Phys.
  Lett. B} {\bfseries 327} (1994) 114--122},
  \href{http://arxiv.org/abs/hep-ph/9312323}{{\ttfamily arXiv:hep-ph/9312323}}.

\bibitem{Larin:1993ju}
S.~A. Larin, T.~van Ritbergen, and J.~A.~M. Vermaseren, {\it {The alpha(s)**3
  correction to Gamma (Z0 ---\ensuremath{>} hadrons)}},
  \href{http://dx.doi.org/10.1016/0370-2693(94)90840-0}{{\em Phys. Lett. B}
  {\bfseries 320} (1994) 159--164},
  \href{http://arxiv.org/abs/hep-ph/9310378}{{\ttfamily arXiv:hep-ph/9310378}}.

\bibitem{Larin:1994va}
S.~A. Larin, T.~van Ritbergen, and J.~A.~M. Vermaseren, {\it {The Large quark
  mass expansion of Gamma (Z0 ---\ensuremath{>} hadrons) and Gamma (tau-
  ---\ensuremath{>} tau-neutrino + hadrons) in the order alpha-s**3}},
  \href{http://dx.doi.org/10.1016/0550-3213(94)00574-X}{{\em Nucl. Phys. B}
  {\bfseries 438} (1995) 278--306},
  \href{http://arxiv.org/abs/hep-ph/9411260}{{\ttfamily arXiv:hep-ph/9411260}}.

\bibitem{Chen:2021rft}
L.~Chen, M.~Czakon, and M.~Niggetiedt, {\it {The complete singlet contribution
  to the massless quark form factor at three loops in QCD}},
  \href{http://arxiv.org/abs/2109.01917}{{\ttfamily arXiv:2109.01917
  [hep-ph]}}.

\bibitem{Matiounine:1998re}
Y.~Matiounine, J.~Smith, and W.~L. van Neerven, {\it {Two loop operator matrix
  elements calculated up to finite terms for polarized deep inelastic lepton -
  hadron scattering}},
  \href{http://dx.doi.org/10.1103/PhysRevD.58.076002}{{\em Phys. Rev. D}
  {\bfseries 58} (1998) 076002},
  \href{http://arxiv.org/abs/hep-ph/9803439}{{\ttfamily arXiv:hep-ph/9803439}}.

\bibitem{Kodaira:1998jn}
J.~Kodaira and K.~Tanaka, {\it {Polarized structure functions in QCD}},
  \href{http://dx.doi.org/10.1143/PTP.101.191}{{\em Prog. Theor. Phys.}
  {\bfseries 101} (1999) 191--242},
  \href{http://arxiv.org/abs/hep-ph/9812449}{{\ttfamily arXiv:hep-ph/9812449}}.

\bibitem{Vogt:2008yw}
A.~Vogt, S.~Moch, M.~Rogal, and J.~A.~M. Vermaseren, {\it {Towards the NNLO
  evolution of polarised parton distributions}},
  \href{http://dx.doi.org/10.1016/j.nuclphysbps.2008.09.097}{{\em Nucl. Phys. B
  Proc. Suppl.} {\bfseries 183} (2008) 155--161},
  \href{http://arxiv.org/abs/0807.1238}{{\ttfamily arXiv:0807.1238 [hep-ph]}}.

\bibitem{Moch:2014sna}
S.~Moch, J.~A.~M. Vermaseren, and A.~Vogt, {\it {The Three-Loop Splitting
  Functions in QCD: The Helicity-Dependent Case}},
  \href{http://dx.doi.org/10.1016/j.nuclphysb.2014.10.016}{{\em Nucl. Phys. B}
  {\bfseries 889} (2014) 351--400},
  \href{http://arxiv.org/abs/1409.5131}{{\ttfamily arXiv:1409.5131 [hep-ph]}}.

\bibitem{Behring:2019tus}
A.~Behring, J.~Bl\"umlein, A.~De~Freitas, A.~Goedicke, S.~Klein, A.~von
  Manteuffel, C.~Schneider, and K.~Sch\"onwald, {\it {The Polarized Three-Loop
  Anomalous Dimensions from On-Shell Massive Operator Matrix Elements}},
  \href{http://dx.doi.org/10.1016/j.nuclphysb.2019.114753}{{\em Nucl. Phys. B}
  {\bfseries 948} (2019) 114753},
  \href{http://arxiv.org/abs/1908.03779}{{\ttfamily arXiv:1908.03779
  [hep-ph]}}.

\bibitem{Tarasov:2020cwl}
A.~Tarasov and R.~Venugopalan, {\it {Role of the chiral anomaly in polarized
  deeply inelastic scattering: Finding the triangle graph inside the box
  diagram in Bjorken and Regge asymptotics}},
  \href{http://dx.doi.org/10.1103/PhysRevD.102.114022}{{\em Phys. Rev. D}
  {\bfseries 102} no.~11, (2020) 114022},
  \href{http://arxiv.org/abs/2008.08104}{{\ttfamily arXiv:2008.08104
  [hep-ph]}}.

\bibitem{Tarasov:2021yll}
A.~Tarasov and R.~Venugopalan, {\it {The role of the chiral anomaly in
  polarized deeply inelastic scattering II: Topological screening and
  transitions from emergent axion-like dynamics}},
  \href{http://arxiv.org/abs/2109.10370}{{\ttfamily arXiv:2109.10370
  [hep-ph]}}.

\bibitem{Blumlein:2021ryt}
J.~Bl\"umlein, P.~Marquard, C.~Schneider, and K.~Sch\"onwald, {\it {The
  three-loop polarized singlet anomalous dimensions from off-shell operator
  matrix elements}},  \href{http://arxiv.org/abs/2111.12401}{{\ttfamily
  arXiv:2111.12401 [hep-ph]}}.

\bibitem{Ahmed:2021spj}
T.~Ahmed, L.~Chen, and M.~Czakon, {\it {Renormalization of the flavor-singlet
  axial-vector current and its anomaly in dimensional regularization}},
  \href{http://dx.doi.org/10.1007/JHEP05(2021)087}{{\em JHEP} {\bfseries 05}
  (2021) 087}, \href{http://arxiv.org/abs/2101.09479}{{\ttfamily
  arXiv:2101.09479 [hep-ph]}}.

\bibitem{Adler:1969er}
S.~L. Adler and W.~A. Bardeen, {\it {Absence of higher order corrections in the
  anomalous axial vector divergence equation}},
\href{http://dx.doi.org/10.1103/PhysRev.182.1517}{{\em Phys. Rev.} {\bfseries
  182} (1969) 1517--1536}.

\bibitem{Gorishnii:1985xm}
S.~G. Gorishnii and S.~A. Larin, {\it {{QCD} Corrections to the Parton Model
  Rules for Structure Functions of Deep Inelastic Scattering}},
  \href{http://dx.doi.org/10.1016/0370-2693(86)90226-1}{{\em Phys. Lett. B}
  {\bfseries 172} (1986) 109--112}.

\bibitem{Ruijl:2017cxj}
B.~Ruijl, T.~Ueda, and J.~Vermaseren, {\it {Forcer, a FORM program for the
  parametric reduction of four-loop massless propagator diagrams}},
  \href{http://dx.doi.org/10.1016/j.cpc.2020.107198}{{\em Comput. Phys.
  Commun.} {\bfseries 253} (2020) 107198},
  \href{http://arxiv.org/abs/1704.06650}{{\ttfamily arXiv:1704.06650
  [hep-ph]}}.

\bibitem{Baikov:2010hf}
P.~Baikov and K.~Chetyrkin, {\it {Four Loop Massless Propagators: An Algebraic
  Evaluation of All Master Integrals}},
  \href{http://dx.doi.org/10.1016/j.nuclphysb.2010.05.004}{{\em Nucl. Phys. B}
  {\bfseries 837} (2010) 186--220},
  \href{http://arxiv.org/abs/1004.1153}{{\ttfamily arXiv:1004.1153 [hep-ph]}}.

\bibitem{Lee:2011jt}
R.~Lee, A.~Smirnov, and V.~Smirnov, {\it {Master Integrals for Four-Loop
  Massless Propagators up to Transcendentality Weight Twelve}},
  \href{http://dx.doi.org/10.1016/j.nuclphysb.2011.11.005}{{\em Nucl. Phys. B}
  {\bfseries 856} (2012) 95--110},
  \href{http://arxiv.org/abs/1108.0732}{{\ttfamily arXiv:1108.0732 [hep-th]}}.

\bibitem{Zijlstra:1992kj}
E.~B. Zijlstra and W.~L. van Neerven, {\it {${\cal O}( \alpha_s^2)$ correction
  to the structure function $F_3$ ($x, Q^2$) in deep inelastic neutrino -
  hadron scattering}},
\href{http://dx.doi.org/10.1016/0370-2693(92)91277-G}{{\em Phys. Lett.}
  {\bfseries B297} (1992) 377--384}.

\bibitem{Trueman:1979en}
T.~L. Trueman, {\it {Chiral Symmetry in Perturbative {QCD}}},
\href{http://dx.doi.org/10.1016/0370-2693(79)90480-5}{{\em Phys. Lett.}
  {\bfseries 88B} (1979) 331--334}.

\bibitem{Breitenlohner:1983pi}
P.~Breitenlohner, D.~Maison, and K.~S. Stelle, {\it {Anomalous Dimensions and
  the Adler-bardeen Theorem in Supersymmetric {Yang-Mills} Theories}},
\href{http://dx.doi.org/10.1016/0370-2693(84)90985-7}{{\em Phys. Lett.}
  {\bfseries 134B} (1984) 63--66}.

\bibitem{Luscher:2021bog}
M.~L\"uscher and P.~Weisz, {\it {Renormalization of the topological charge
  density in QCD with dimensional regularization}},
  \href{http://arxiv.org/abs/2103.15440}{{\ttfamily arXiv:2103.15440
  [hep-ph]}}.

\bibitem{Ju:2021lah}
W.-L. Ju and M.~Sch\"onherr, {\it {The q$_{T}$ and
  \ensuremath{\Delta}\ensuremath{\phi} spectra in W and Z production at the LHC
  at N$^{3}$LL'+N$^{2}$LO}},
  \href{http://dx.doi.org/10.1007/JHEP10(2021)088}{{\em JHEP} {\bfseries 10}
  (2021) 088}, \href{http://arxiv.org/abs/2106.11260}{{\ttfamily
  arXiv:2106.11260 [hep-ph]}}.

\bibitem{Bos:1992nd}
M.~Bos, {\it {Explicit calculation of the renormalized singlet axial anomaly}},
   \href{http://dx.doi.org/10.1016/0550-3213(93)90479-9}{{\em Nucl. Phys. B}
  {\bfseries 404} (1993) 215--244},
  \href{http://arxiv.org/abs/hep-ph/9211319}{{\ttfamily arXiv:hep-ph/9211319}}.

\bibitem{Zoller:2013ixa}
M.~Zoller, {\it {OPE of the pseudoscalar gluonium correlator in massless QCD to
  three-loop order}},  \href{http://dx.doi.org/10.1007/JHEP07(2013)040}{{\em
  JHEP} {\bfseries 07} (2013) 040},
  \href{http://arxiv.org/abs/1304.2232}{{\ttfamily arXiv:1304.2232 [hep-ph]}}.

\bibitem{Ahmed:2015qpa}
T.~Ahmed, T.~Gehrmann, P.~Mathews, N.~Rana, and V.~Ravindran, {\it
  {Pseudo-scalar Form Factors at Three Loops in QCD}},
  \href{http://dx.doi.org/10.1007/JHEP11(2015)169}{{\em JHEP} {\bfseries 11}
  (2015) 169},
\href{http://arxiv.org/abs/1510.01715}{{\ttfamily arXiv:1510.01715 [hep-ph]}}.

\end{thebibliography}\endgroup
\bibliographystyle{utphysM}
\end{document}